# Evaluating Wireless Reactive Routing Protocols with Linear Programming Model for Wireless Ad-hoc Networks

N. Javaid[1], M. Ilahi[1], R. D. Khan[2], L. Ali[1], U. Qasim[3], Z. A. Khan[4]

[1,2]COMSATS Institute of Information Technology, [1]Islamabad, [2]Wah Cant, Pakistan.
[3]University of Alberta, Alberta, Canada.
[4]Faculty of Engineering, Dalhousie University, Halifax, Canada.

## ABSTRACT

InWireless Ad-hoc Networks, nodes are free to move randomly and organize themselves arbitrarily, thus topology may change quickly and capriciously. In Mobile Ad-hoc NETworks, specially Wireless Multi-hop Networks provide users with facility of quick communication. In Wireless Multi-hop Networks, routing protocols with energy efficient and delay reduction techniques are needed to fulfill users' demands. In this paper, we present Linear Programming models to assess and enhance reactive routing protocols. To practically examine constraints of respective Linear Programming models over reactive protocols, we select AODV, DSR and DYMO. It is deduced from analytical simulations of Linear Programming models in MATLAB that quick route repair reduces routing latency and optimizations of retransmission attempts results efficient energy utilization. To provide quick repair, we enhance AODV and DSR. To practically examine the efficiency of enhanced protocols in different scenarios of Wireless Multi-hop Networks, we conduct simulations using NS-2. From simulation results, enhanced DSR and AODV achieve efficient output by optimizing routing latencies and routing load in terms of retransmission attempts.

## I. INTRODUCTION

In Wireless Multi-hop Networks (WMhNs), links frequently change due to wireless nature. Routing protocols are used to provide accurate routes. The protocols are divided into two main categories based on their routing operations to accurately discover and compute routes; reactive and proactive. Protocols belong to the former category calculate and make available route(s) when data demand arrives, whereas, the protocols in the later category calculate routes periodically and are independent from data demands.

In [1], authors evaluate AODV [2][3] and DSR [4][5] with respect to the varying number of Constant Bit Rate (CBR) resources. The authors in [6], evaluate performance of DSR and AODV with varying number of sources (10 to 40 sources with different pause times). Problem from a different perspective in [7], using a simulation model with a dynamic network size and is practically examined for Destination-Sequence Distance Vector (DSDV) [8], AODV, DSR and Temporally-Ordered Routing Algorithm (TORA) [9].

In WMhNs, reactive protocols are responsible to find accurate routes and provide quick repair after detecting route breakages. This work is devoted to study routing capabilities of three reactive protocols named as AODV, DSR and DYnamic MANET On-demand (DYMO) [10] in different network cases of WMhNs. The contribution of this work includes: (i) construction of LP_model for WMhNs requirements and analytical simulations of the models for selected protocols, (ii) enhancements in AODV and DSR, (iii) performance evaluation of the selected routing protocols with respect to framework of network constraints (iv) analytical analysis of mobility, traffic rates and scalability properties of the selected routing protocols using NS-2.

## II.     RELATED WORK AND MOTIVATION

The authors in [11], examine performance of proactive routing protocols. They set up a mathematical model to optimize proactive routing as well as balance the routing overhead of the protocols between routing accuracy. Their model is generalized for HELLO intervals and they deduce that by optimizing the time interval of a HELLO message, the proactive routing protocol will have less routing overhead and high delivery rate. To evaluate routing overhead, their mathematical model is generalized for proactive class while in our work, we discuss the behaviour of reactive protocols.

The study about mobility impacts on performance of reactive protocols is presented by authors [12]. They examine how statistics of the path durations including probability density functions vary with respect to the parameters such as mobility model, relative speed, number of hops, and radio range. They also model a framework for mobility constraints and communication traffic patterns for approximation of the path duration distribution.

Contrary to the above mentioned works, in this paper, the novel contribution is construction of the mathematical framework to study reactive routing protocols for WMhNs. For this purpose, we develop *LP_models* that list all possible constraints for different mobilities and varying network flows. In this framework, throughput, energy cost in terms of routing packets and delay is objective functions. We further enhance DSR by proposing a quick route repairing method and simulate AODV with and without link layer feedback. Which protocol gives an optimal solution in what scenario by satisfying *LP_model* constraints is discussed in detail by practically evaluating them in NS-2.

## III.     PROBLEM FORMULATION USING LINEAR PROGRAMMING

We formulate LP_models for performance metrics; throughput *(max T_avg)*, energy cost *(min CE)*, and time cost (min CT), for WMhNs. These models are discussed below in detail:

- Lp_Model for Maximizing Throughput (Max t_avg)

A protocol is aimed to provide efficient data delivery by end-to-end path calculations. These parameters along with their effects on the objective function *(max T_avg)* are discussed below:

$dr$ denotes an individual data request in a set of all data requests, *DR*, such that $dr \in DR$.

$\tau$ and $T$ specify unit time and simulation time, respectively, where .

$Rec_{dr}$ is number of successfully received data packet(s). Only $Rec_{dr}$ is considered for throughput measurements.

Routing protocols are supposed to provide accurate routes for each $dr$. $p_{nr}$ represents the probability of no route available for $dr$ during route discovery process.

For each $dr$, a reactive protocol, $rp$, performs route discovery to find the requested destination. The control packets $c_p^{rp}$ generated by a reactive routing protocol for route discovery rates are $\alpha^{rd}$.

When end-to-end path has been calculated then $\alpha^{dt}$ data transmission rate corresponds to data request arrival from sender side. $d_p^{rp}$ is the number of data packets impeded in the control informational message of a routing protocol currently transmitted on a channel from source node.

let $\alpha$ is rate parameter, thus, data request arrival(s) and successfully received data packets rates are represented by $\alpha_{tra}$ and $\alpha_{rec}$, respectively. Generally, $\alpha_{tra}$ is the data request transmission rate by the source node, while $\alpha_{rec}$ is the rate of received data packets rates at destination node.

$\beta_{avail}$ denotes available bandwidth value of a channel during $\tau$.

In wireless communication, links among nodes are frequently changed. In $lb \in LB$, the object $lb$ represent the link breakage rate at any instant time $\tau$, and $LB$ symbolizes the whole link breakage rate during all the network connectivity period ($T$).

A network connectivity graph is represented as $G(V,E)$; here $V$ are the vertices and $E$ represents edges or links between the nodes. Any two nodes which are within the maximum allowable transmission rang $R_{(i,j)}^{max}$ can directly communicate i.e., it is necessary the difference of distance between the upstream and downstream links is less than or equal to $R_{(i,j)}^{max}$ for a node pair in a connected network. $LC_{max}$ is the maximum number of link changes value during connectivity period of a network.

$\alpha_{lr}^{rp}$ is the link repair response rate produced by a routing protocol correspond to each $r$.

$rn$ represents a node in a route among a set of all active routes; $rn \in RN$.

$p_{s\_RD}$ and $p_{s\_RM}$ denote probability of successful Route Discovery (RD) and Route Maintenance (RM), respectively. These two processes are involved in reactive routing protocols.

Here, we are considering only received packets for throughput measurements. Thus, objective function $max\,T_{avg}$, is expressed as:

$$maxT_{avg} = \frac{\sum_{dr \in DR}(1-p_{nr})Rec_{dr}}{\sum_{\tau \in T}}(\frac{B}{s})\ dr \in DR \qquad (1)$$

**Subject to:**

$$\forall l \in L \qquad c_p^{rea} + d_p^{rea} \leq \beta_{avail} \qquad c_p \in C_P \qquad (1.a)$$
$$\forall lb \in LB \qquad sum_{r \in R}\alpha_{lr} \leq LC_{max} \qquad (1.b)$$
$$\forall rn \in RN \qquad dist_j - dist_i \in R_{(i,j)}^{max} \qquad \forall(i,j) \in E \qquad (1.c)$$
$$p_{s\_RD} \leq 1 \qquad \forall dr \in DR \qquad (1.d)$$
$$p_{s\_RM} \leq 1 \qquad \forall br \in BR \qquad (1.e)$$

- LP_MODEL FOR ROUTING DELAY CT

Routing delay, $CT$, is time required by a reactive protocol for processing incoming $dr$. RD and RM are two processes which effect time costs.

For each $dr$, the term time cost, $CT$, is used for routing messages. During RD and RM processes, the cost is represented by the terms $CT_{RD}$, $CT_{RM}$, respectively.

$\tau_{RD}^{rp}$ specifies the route (re)discoveries time required for a reactive routing protocol in

response to a single RREQ.

$\tau_{RM}^{rp}$ specifies the link monitoring and repairing time in RM activity associated with a reactive protocol in response to $dr$.

$\tau_{cri}$ stipulates critical delay value, which means that remaining is not enough to further transmission for $dr$. Such a situation arises in case of delay in the route discovery in dense network, high data rates and high mobilities due to extensive link breakage. All these situations can result buffer_time_out, and as a result deletes the requested RREQ.

Let $CT$ be the required minimizing objective function used to express routing delay generated by reactive routing protocols, we write this as:

$$min\ CT \qquad (2)$$

**Subject to**

$$\forall dr \in DR \quad CT_{RD}^{rp} < \tau_{cri} \qquad (2.a)$$
$$\forall dr \in DR \quad CT_{RM}^{rp} < \tau_{cri} \qquad (2.b)$$

- LP_MODELING FOR ROUTING OVERHEAD CE

The parameters along with their effects on objective function ($min\ CE$) are discussed below:

• Routing overhead, $CE$, represents the number of routing packets produced by a routing protocol and it depends upon the nature and operations of protocols.

For each RREQ $dr$, the general term energy cost; $CE$ is used for routing messages, during RD, it is so called $CE_{RD}$ and for RM process it is represented by $CE_{RM}$.

$\alpha_{RD}^{rp}$ indicates the route (re)discoveries rate associated with a reactive routing protocol in response to $dr$. $CE_{RD}^{rp}$ is the number of control packets produced during route (re)discoveries.

$\alpha_{RM}^{rp}$ specifies the route maintenance rate associated with a reactive routing protocol in response to $dr$. $CE_{RM}^{rp}$ is the number of control packets produced collectively during link monitoring and repairing process.

$\beta_{cri}$ stipulates critical bandwidth which restricts further transmission for $dr$ data. Such a situation arises in case of high data rates and mobilities.

Let $min\ CE$ is the minimizing objective function used to express routing overhead generated by reactive protocols. We can write this as:

$$min\ CE \qquad (3)$$

**Subject to**

$$\forall dr \in DR \quad \alpha_{RD}^{rp} CE_{RD}^{rp} < \beta_{cri} \qquad (3.a)$$

$$\forall dr \in DR \quad \alpha_{RM}^{rp} CE_{RM}^{rp} < \beta_{cri} \qquad (3.b)$$

As, we discussed above that for analyzing the effect of network constraints, we select AODV, DSR and DYMO. The basic operations of the protocols are discussed below:

## IV. REACTIVE PROTOCOLS WITH THEIR BASIC OPERATIONS

The protocols use two basic operations; RD and RM. The total Energy Cost ($CE$) for reactive protocol, $rp$; $CE_{total}^{rp}$ [14]:

$$CE_{total}^{rp} = CE_{RD}^{rp} + CE_{RM}^{rp} \qquad (4)$$

where, $CE_{RD}^{rp}$ and $CE_{RM}^{rp}$ represent energy cost for RD and RM processes, respectively.

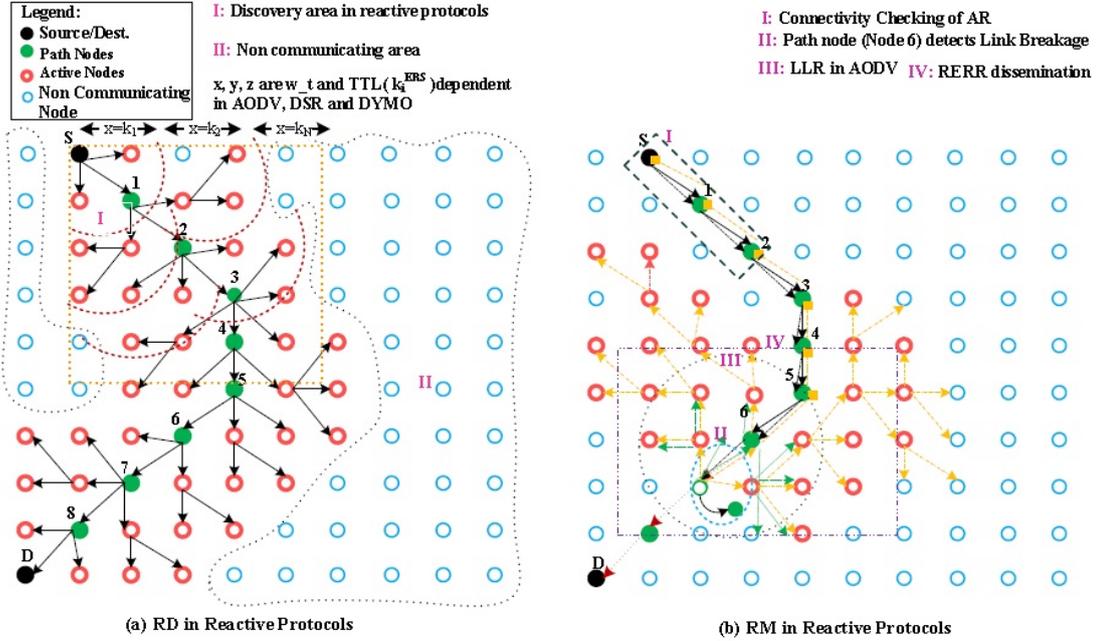

Fig. 1. Basic routing operations in reactive protocols

- $CE_{RD}^{rp}$

Expanding Ring Search (ERS) [14][15] is used as optimization techniques in AODV, DSR and DYMO during RD. In ERS, flooding is controlled by Time-To-Live (TTL) values to limit the broadcast. A source node $S_n$ may receive RREPs from the nodes that contain alternate (short) route for the desired destination $D_n$, as shown in Fig. 1 (a). $S_n$ establishes a path to $D_n$ which contains $9$ hops. These replies are only used in AODV and DSR and are known as gratuitous RREPs (*grat. RREPs*). The destination RREPs are generated by the $D_n$ (destination RREPs are generated in all the three reactive protocols). So, control packet cost for RD; $CE_{RD}^{rp}$ can be calculated as [13]:

$$CE_{RD}^{rp} = \sum_{i=1}^{M} CE_k(i) \qquad (5)$$

Let, $M$ is number for maximum rings during $RD$. The generation of RREP(s) in AODV and DSR is also due to the valid routes in Routing Table $(RT)$ or in Route Cache $(RC)$, so, $M$ for DSR and AODV can be less than DYMO, because of absence of grat. RREPs in DYMO. Let $d_{avg}$ is the average degree of nodes. The cost of any $k(i)$ can be calculated as $CE_k(i)$:

$$CE_k(i) = d_{avg} + d_{avg} \sum_{i=1}^{N_k} i \qquad (6)$$

Where, $N_k$ represents the total number of nodes in the ring $k_i$.

- $CE_{RM}^{rp}$

In $RM$ process, different protocols pay different costs for link monitoring and also there are different costs for different supplementary maintenance strategies in case of link breakages. DYMO and AODV generate HELLO messages to check the connectivity of RN, while DSR gets the link level feedback from link layer.

In DYMO, link breakages in networks cause broadcasting of RERR messages. When the probability of unsuccessful local link repair ($LLR$) and is represented by; $p_{us}^{llr}$ leads to the dissemination of RERRs in AODV, as shown in Fig. 1.b, block (III). On the other hand, DSR piggy-backs RERR messages along with next RREQs in the case of route re-discovery process [15], while these RERR messages are generated in the case of success of $PS$, as depicted in Fig. 1(b) block I.

In AODV, after unsuccessful $RD$ and after detecting link breakage in DYMO, RERR messages are broadcasted by the node which detects any link break and route rediscovery process is started through source node.

$$CE_{RM}^{AODV} = CE_{HELLO} + |sgn\ lb_{RN}| \sum_{i=1}^{N_{llr}} i$$
$$+ |sgn\ p_{us}^{llr}| + \sum_{N_{rerr}} i \qquad (7)$$

We also compare the performance of AODV without HELLO messages and refer it as AODV-LL, and in this case link layer feed back is used. As in [4], HELLO_INTERVAL for AODV is $1s$, and ALLOWED_HELLO_LOSS is $2$, thus a link can be considered as broken after expiration of ALLOWED_HELLO_LOSS value. To increase the efficiency of AODV, quick detection of link breakage is needed. The energy cost for AODV-LL is given as:

$$CE_{RM}^{AODV-LL} = |sgn\ lb_{AR}| \sum_{i=1}^{N_{llr}} i + |sgn\ p_{us}^{llr}| + \sum_{N_{rerr}} i \qquad (8)$$

Whereas, CE$_{HELLO}$ is the energy cost of HELLO messages for link monitoring. Further details for this cost is available in [9]. When link breakages of RN ($lb_{RN}$) occurs that cause initialization of $LLR$. After unsuccessful $LLR$, RERR messages are broadcasted in AODV, and $N_{rerr}$ represents the number of nodes that receives RERR messages.

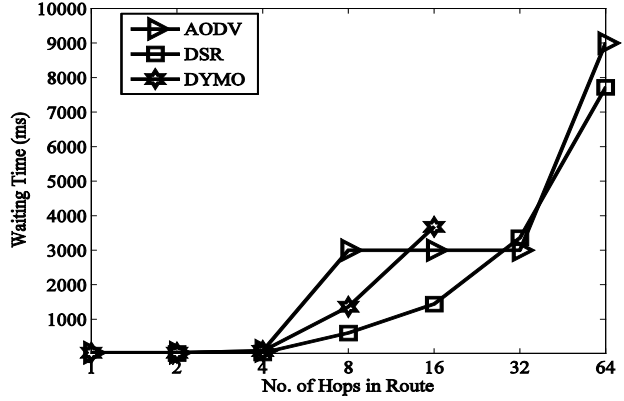

Fig.2. Delay in Route Discovery

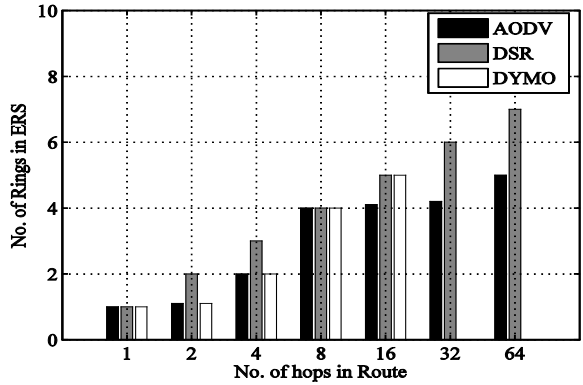

Fig.3. Routing Overhead in Route Discovery

$$CE_{HELLO} = \frac{\tau_{route\_in\_use}}{\tau_{H\_interval}} \times N_{RN} \qquad (9)$$

Let $\tau_{route\_in\_use}$ is the total time in which route remains in use, while $\tau_{H_interval}$ specifies the $HELLO\_INTERVAL$ (which is 1s in AODV). Moreover, $N_{RN}$ represent the number of nodes in Active Routes (RN(s)).

Like AODV, in case of DSR's PS technique can reduce both the energy and time cost to be paid by a reactive protocol by diminishing the route re-discovery. In the case of successful PS, RERR messages are broadcasted to neighbors for the deletion of useless routes. Whereas, the absence of alternate route(s) in RC leads to the failure of PS. In this situation, RERR messages are to be sent by piggy-backing them in the next RREQ messages during RD process.

$$CE_{RM}^{DSR} = \sum_{i=1}^{n_{ps}} i \qquad (10)$$

Where, $n_{ps}$ denote the node that salvage the packet successfully.

$$CE_{RM}^{DYMO} = CE_{HELLO} + |sgn\ lb_{AR}| \sum_{i=1} N_{rerr} i \qquad (11)$$

We analytically simulate LP_model for $CE$ and $CT$ of selected protocols in MATLAB. In Fig. 2, respective $waiting\_time$ of the protocols is depicted. AODV and DSR need efficient mechanism to reduce $CT$ (refer Fig. 2). We therefore enhance AODV and DSR in order to provide quick RD and RM. From Fig. 3, it is clear that routing overhead of DSR is high because of the high value of $M$. To reduce this value, an efficient packet salvaging and route caching for routing is required. Enhance DSR (DSR-M) provides more accuracy and efficiency.

## V. SIMULATION RESULTS

We evaluate performance of the proposed framework in NS-2. For simulation setup, we choose Random Way point mobility model. We take mobilities and traffic flows scenarios for our evaluation. The area specified is 1000m×1000m field presenting a square space to allow mobile nodes to move inside. All of the nodes are provided with wireless links of a bandwidth of 2Mbps to transmit on. Simulations are run for 900s each. For evaluating mobilities effects, we vary pause time from 0s to 900s for 50 nodes within two different speeds of 2m/s and 303/s separately. For evaluating different network flows with 15m/s speed and fixed pause time of 2s, 1) different scalabilities from 10 to 100 nodes 2) traffic rate of 2, 4,8. 16 and 32 packs/s for 50 nodes.. We evaluate and compare the protocols by three performance parameters; throughput, CT in terms of average end-to-end delay, and CE in terms of normalized routing load.

- THROUGHPUT

For throughput measurement we consider successfully received data packets, as mentioned in eq. 1.

In high mobilities, constraints in eq. 1.(b)(c)(d) are more critical to be satisfied for throughput. DSR gives high throughput in Fig. 4 because of accurate and efficient mechanisms for RD and RM processes by satisfying constraints 1.c and 1.d due to low speeds of $2m/s$. From Fig. 5, it is depicted that in a very high dynamic situation, RC of DSR becomes ineffective, as, there is no mechanism to delete the stale routes from RC, and RERR messages are disseminated not traditionally as in other protocols as in eq. 10; thus, the protocol fails to converge at this mobility speed, that is why effected specially by constraints in eq. 1.(c)(d). While AODV checks the route with valid time and avoids using the invalid routes from RT, thus, achieves more successful probability of RD (as in constraint eq. 1.(c)).

In AODV-LL, quick detection and retirement make this protocol more efficient than AODV, and DSR-M reduces generation of stale route information by reducing TAPE_CACHE_SIZE (In DSR-M, we change TAPE_CACHE_SIZE from 1024 to 256, this modification results quick updating of RC). Moreover, the HELLO messages and LLR (as mentioned in eq. 8) make able the protocol to handle highest rate of mobility and fulfill eq. 1.(b)(d) constraints, thus, overall converges in dynamic situations. The bad behaviour of DYMO among reactive protocols in response to mobility by showing overall less throughput value as is noticed in Fig. 45. The absence of any supplementary mechanism make its performance low against respective constraints of throughput in high mobilities (in eq. 11 only generations of RERR messages despite of initiating any repairing mechanism).

Conducted simulation results from Fig. 6 and 7, AODV shows convergence for all data rates and all scalability, whereas DSR is less scalable while DYMO degrades its performance in more population of nodes. In [4], it is specified that " AODV can better handle a wireless network of tens to thousand nodes ", therefore, it performs better among reactive protocols for high network flows. The presence of grat. RREPs and time-based routing activities that makes able the protocol to perform well by always choosing a fresher end-to-end path. The route deletion using RERR messages is also traditional and disseminates quick information after failure of $LLR$ as mentioned in eq. 7. It also maintains the predecessor list; RERR packets reach all nodes using a failed link on its route to any desired destination.

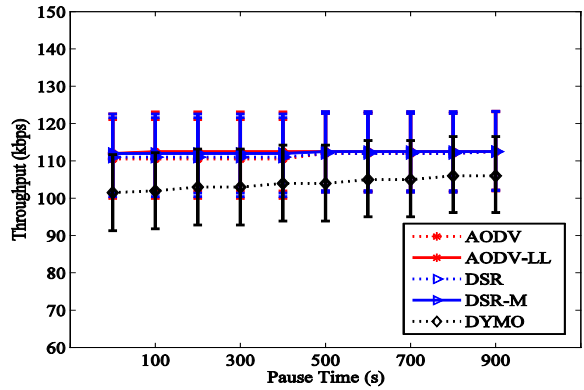

*Fig.4. Throughput of Reactive Protocols at 2m/sec Mobility*

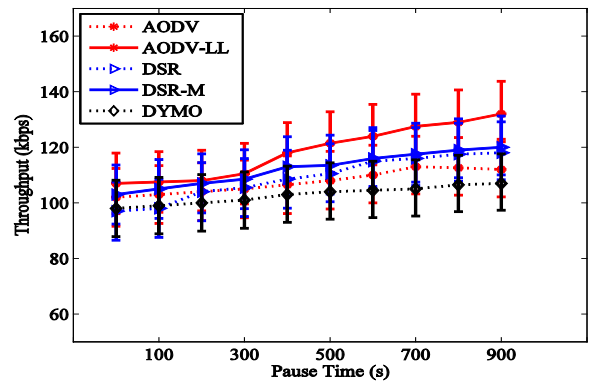

Fig.5. Throughput of Reactive Protocols at 30m/sec Mobility

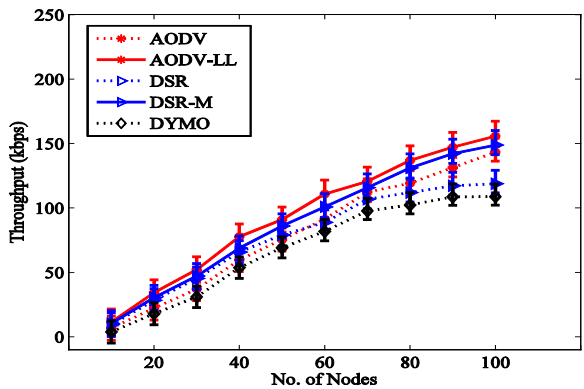

Fig. 6. Throughput of Reactive Protocols vs Scalability

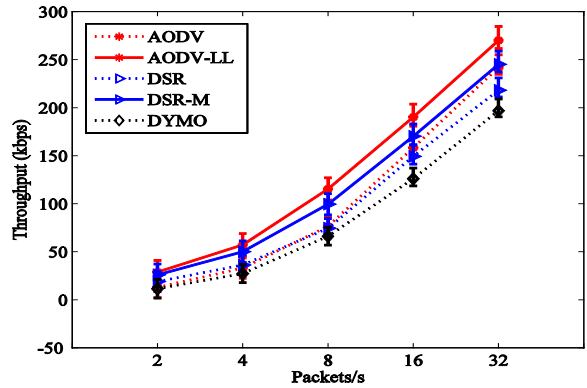

Fig. 7. Throughput of Reactive Protocols vs Traffic Rate

- COST OF TIME

AODV among reactive protocols in attains the highest delay. Because in local repair for link breaks in routes sometimes results in increased path lengths (against eq. 2.a constraint). In AODV-LL, E2ED becomes much less as compared to routing latency of AODV, because LLR initiation and repairement starts quickly after receiving link layer feed-back (link layer beacon messages to check the connectivity is send $100$ times in a second, and after $8$ connective failure notify link breakage), as depicted in Figs. 8,9,10 and 11. DSR does not implement $LLR$ [10],[11], therefore, its $CT$ value is less than AODV but during moderate and high mobility RC search fails frequently and results high routing delay.

At higher mobility, DSR suffers the higher $CT$ value, as portrayed in Fig. 9. The reasons include: for RD, it first searches the desired route in the RC and then starts RD if the search fails, moreover, this searching is also performed during RM for PS process. Therefore, in high mobilities with high speeds, it does not give feasible solution against eq. 2.a and 2.b constraints as shown in Fig. 9. DYMO produces the lowest $CT$ value among reactive protocols because it only uses ERS for route finding which results low delay; as checking the RC (in DSR) and RT (in AODV) before the RD cause delay of node traversal information. DSR-M also gives low value for routing latencies while considering high mobilities and scalabilities (in Figs. 9,10 and 11).

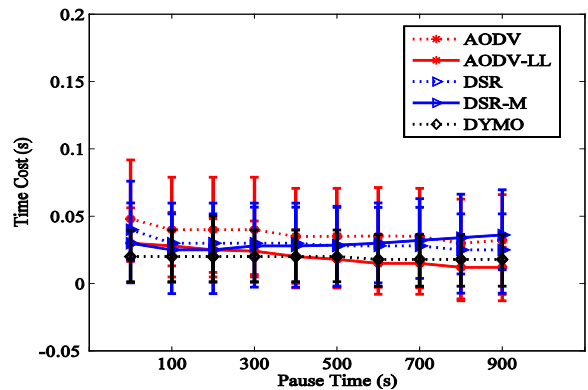

Fig. 8. Time cost analysis of Reactive Protocols at 2m/sec Mobility

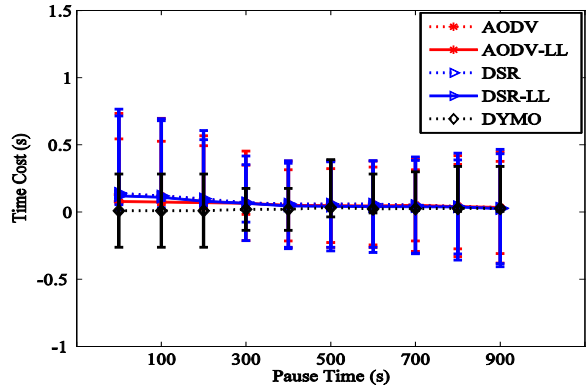

Fig. 9. Time cost analysis of Reactive Protocols at 30m/sec Mobility

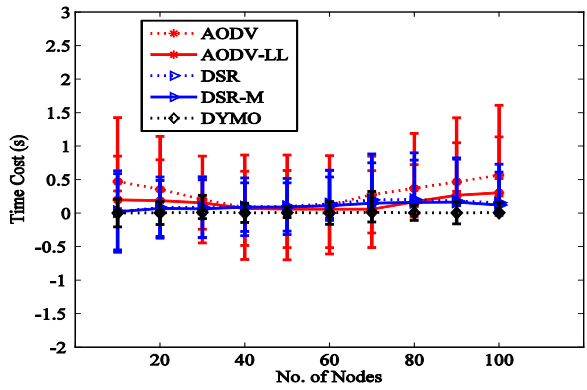

Fig. 10. Time cost analysis of Reactive Protocols vs Scalability

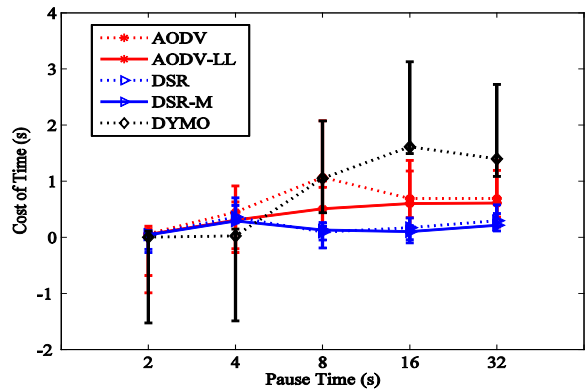

Fig. 11. Time cost analysis of Reactive Protocols vs Traffic Rate

DYMO does not use any supplementary strategies like grat. RREPs, PSing, RCing or LLR, therefore it suffers lowest delay in low traffic while produces high latency in high data rates. On the other hand, absence of the mechanism keeps the lowest delay cost of DYMO in all scalability, as shown in Fig. 9. PS and grat. RREPs keep the delay low in medium and high traffic scenarios for DSR but first checking the RC instead of simple ERS based RD process augments the delay when population increases, thus, more delay of DSR is presented in Fig. 9, as compared to DYMO. AODV experiences the highest E2ED in all scalability due to LLR process (Fig. 9).

- COST OF ENERGY

One common and noticeable behaviour of all reactive protocols in Figs. 12 and 13 is that at high speeds and or high mobilities, energy cost is higher as compared to moderate and low mobilities and or speeds. This is because of more link breakages during high dynamic situations, and all of the on-demand protocols initiate route repairing mechanisms to re-establish broken paths along with dissimilation of RERR messages.

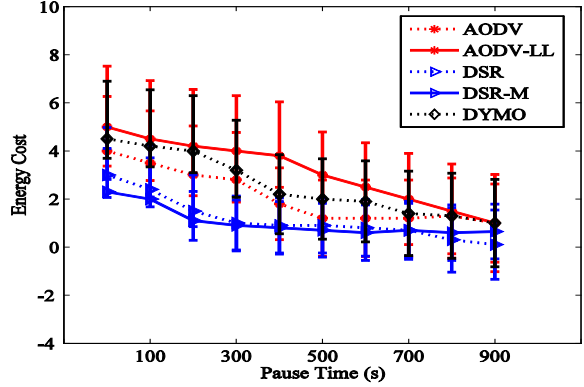

Fig. 12. Energy cost analysis of Reactive Protocols at 2m/sec Mobility

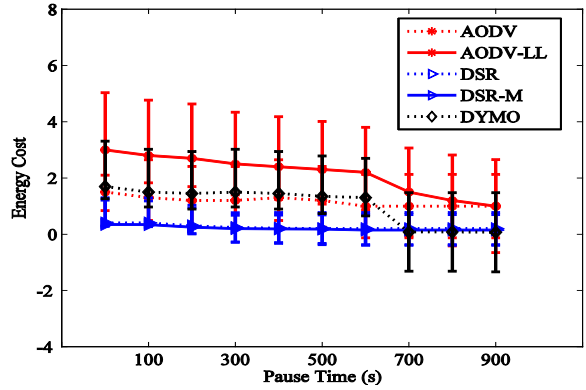

Fig. 13. Energy cost analysis of Reactive Protocols at 30m/sec Mobility

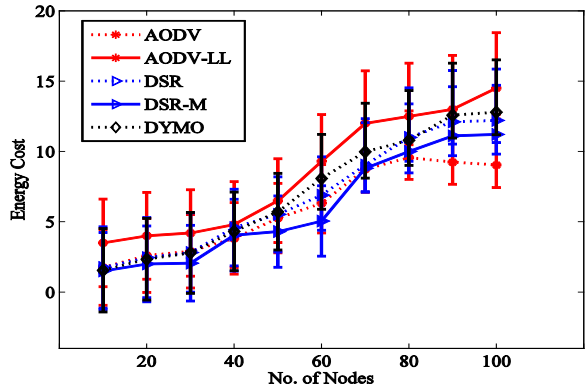

Fig. 14. Energy cost analysis of Reactive Protocols vs Scalability

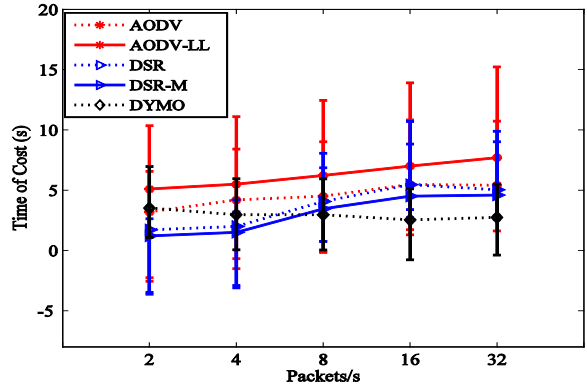

Fig. 15. Energy cost analysis of Reactive Protocols vs Traffic Rate

DSR generates lowest number of control packets compared to rest of two because PSing and RCing reduce the routing overhead of RD and RM, respectively. The highest routing overhead is produced by DYMO because of simple ERS without any optimization technique and re-discovery process for repairement of broken route without any quick repair mechanism make its routing overhead highest (in Figs. 12, 13, 14, and 15) dissemination of RERR messages after unsuccessful RD increase control packets comparative to DSR (eq. 11 comparative to eq. 10).

AODV-LL increase routing overhead by generating more control packets for LLR after quick detection of link breakages. On the other hand, DSR-M gives low NRL value as compared to DSR (Fig. 12 and 13) because of reducing *grat. RREPs* due to small sized RC.

In medium and high densities, routing load of DYMO is less than DSR and AODV, as in Fig. 14 and 15. While in medium and more density, AODV attains the highest routing load. The HELLO messages to check the connectivity of active routes, *LLR* and *grat. RREPs* , increase the generation of control packets. Whereas, PS of DSR along with promiscuous listening mode jointly reduce the routing overhead in low scalabilities. Each node participating in RD process (including intermediate nodes) of DSR, learns the routes to other nodes on the route. PS technique is used to get routes from RC of the intermediate nodes. This strategy is used to quickly access and to solve broken link issues by providing alternative route. However, in large population of nodes, intermediate nodes generating more *grat. RREPs* increase routing overhead. Same as that in AODV-LL in mobilities, in the case of scalabilities, as depicted in Fig. 14.

## VI. CONCLUSION AND FUTURE WORK

WMhNs provide users with facility of quick communication. Different routing protocols are used to facilitate users in high mobilities and scalabilities. Energy efficiency and delay reduction are two important factors to check the performance of a protocol. To evaluate these factors, this paper contributes LP_models for WMhNs. To practically examine the respective constraints over reactive routing protocols, we select AODV, DSR and DYMO. We relate effects of RD and RM strategies of the selected protocols over WMhNs' constraints to check energy efficiency and delay reduction of chosen protocols in different scenarios in NS-2 while considering throughput, cost of time and cost of energy. Quick route repair and optimizations of retransmission attempts result in better performance of the protocol by reducing energy utilization and routing latencies. For quick deletion of stale route entries in DSR, we reduce *TAP_CACHE_SIZE* of DSR (DSR-M) and compare it with original DSR. For quick repairement, we compare AODV with and without link level feed back. Finally we deduce that AODV-LL due to quick repairment produces highest throughput by providing feasible solution for *max $T_{avg}$* and *min CT*.

In future, we are interested to extend this analysis on the issues addressed in [16-20].

### REFERENCES

1. V. Naumov and T. Gross, 2005. Scalability of routing methods in ad hoc networks.

Performance Evaluation, 62(1): pp. 193—209.
2. Perkins, C.E. and Royer, E.M., 1999. Ad-hoc on-demand distance vector routing. Mobile Computing Systems and Applications, 1999. Proceedings. WMCSA'99, pp. 90--100.
3. Perkins, C. and Belding-Royer, E. and Das, S., 2003. Ad hoc On-Demand Distance Vector (AODV) Routing. IETF RFC3561, Available: http://www.ietf.org/rfc/rfc3561.txt.
4. Johnson, D. and Hu, Y. and Maltz, D., 2007. The dynamic source routing protocol (DSR) for mobile ad hoc networks for IPv4. RFC4728, pp. 2—100.
5. Broch, J. and Maltz, D.A. and Johnson, D.B. and Hu, Y.C. and Jetcheva, J., 1998. A performance comparison of multi-hop wireless ad hoc network routing protocols. Proceedings of the 4th annual ACM/IEEE international conference on Mobile computing and networking, pp. 85—97.
6. Hassan, Y.K. and El-Aziz, M.H.A. and El-Radi, A.S.A., 2010. Performance evaluation of mobility speed over MANET routing protocols. International Journal of Network Security, 11(3), pp. 128—138
7. Layuan, L. and Chunlin, L. and Peiyan, Y., 2007. Performance evaluation and simulations of routing protocols in ad hoc networks. Computer Communications, 30(8), pp. 1890—1898.
8. Perkins, C.E. and Bhagwat, P., 1994. Highly dynamic destination-sequenced distance-vector routing (DSDV) for mobile computers. ACM SIGCOMM Computer Communication Review, 24 (4), pp. 234—244.
9. Park, V.D. and Corson, M.S., 1997. A highly adaptive distributed routing algorithm for mobile wireless networks. INFOCOM'97. Sixteenth Annual Joint Conference of the IEEE Computer and Communications Societies, pp. 1405—1413.
10. Chakeres, I.D. and Perkins, C.E., 2008. Dynamic MANET on-demand routing protocol. IETF Internet Draft, draft-ietf-manet-dymo-12. Txt.
11. Thorup, R.E., 2007. Implementing and evaluating the DYMO routing protocol. Thesis: Citeseer.
12. Bai, F. and Sadagopan, N. and Krishnamachari, B. and Helmy, A., 2004. Modeling path duration distributions in MANETs and their impact on reactive routing protocols. Selected Areas in Communications, IEEE Journal on, 22(7), pp. 1357—1373.
13. Javaid, N. and Bibi, A. and Javaid, A. and Malik, S.A., 2011. Modeling routing overhead generated by wireless reactive routing protocols. Communications (APCC), 2011 17th Asia-Pacific Conference on, pp. 631—636.
14. Deng, J. and Zuyev, S., 2008. On search sets of expanding ring search in wireless networks. Ad hoc networks. 6(7), pp. 1168—1181.
15. Iqbal, M.A. and Wang, F. and Xu, X. and Eljack, S.M. and Mohammad, A.H., 2011. Reactive routing evaluation using modified 802.11 a with realistic vehicular mobility. Annals of Telecommunications 66 (11), pp. 643—656.
16. Javaid. N, Javaid. A, Khan. I. A, Djouani. K, "Performance study of ETX based wireless rout- ing metrics," 2nd IEEE International Conference on Computer, Control and Communications (IC4-2009), Karachi, Pakistan, pp.1-7, 2009.
17. Dridi. K, Javaid. N, Djouani. K, Daachi. B, "Performance Study of IEEE802.11e QoS in EDCF- Contention-Based Static and Dynamic Scenarios", 16th IEEE International Conference on Electronics, Circuits, and Systems (ICECS2009), Hammamet, Tunisia, 2009.
18. Dridi. K, Javaid. N, Daachi. B, Djouani. K, "IEEE 802.11 e-EDCF evaluation through MAC-layer metrics over QoS-aware mobility constraints", 7th International Conference on Advances in Mobile Computing & Multimedia (MoMM2009), Kuala Lumpur, Malaysia, 2009.
19. Javaid. N, Bibi, A, Djouani, K., "Interference and bandwidth adjusted ETX in wireless multi-hop networks", IEEE International Workshop on Towards Samart Communications and Network Technologies applied on Autonomous Systems (SaCoNaS2010) in conjunction with 53rd IEEE International Conference on Communications (ICC2010), Ottawa, Canada, 2010., Miami, USA, 1638-1643, 2010.
20. Javaid. N, Bibi, A, Djouani, K., "Interference and bandwidth adjusted ETX in wireless multi-hop networks", IEEE International Workshop on Towards Samart Communications and